\documentclass[preprint, 12pt]{aastex} 
\singlespace
  
\begin{document} 

\title{Stochastic Fermi Acceleration of sub-Relativistic Electrons and Its Role in Impulsive Solar Flares} 
\author {Robert Selkowitz and Eric G. Blackman}
\affil{Dept. of Physics and Astronomy, and Laboratory for Laser Energetics, 
University of Rochester, Rochester NY, 14627}
\begin{abstract}

We reexamine stochastic Fermi acceleration (STFA) in the low energy
(Newtonian) regime in the context of solar flares.  The particle energization
rate depends  a dispersive term  and a coherent gain term.
The energy dependence of pitch angle scattering  is important for
determining the electron energy spectrum.  For scattering by whistler wave
turbulence, STFA produces a quasi-thermal spectrum. A second
well-constrained scattering mechanism is needed for STFA
to match the observed $10-100$keV non-thermal spectrum. We suggest
that STFA most plausibly acts as phase one  of a two
phase particle acceleration engine in impulsive flares:
STFA can match the thermal spectrum below $10$kev, and possibly the
power law spectrum between $10$ and $100$keV, given the proper
pitch angle scattering. However, a second phase,
such as shock acceleration  at loop tops, is likely required to match the
spectrum above the observed knee at $100$keV. Understanding this
knee, if it survives further observations, is tricky.

\bigskip
\bigskip
\end{abstract}


\section{Introduction}

Fermi acceleration was first proposed as a mechanism for cosmic ray 
acceleration \citep{fermi49,Fermi}.  In the original model, compressive perturbations 
in the Galactic magnetic field, associated with molecular clouds, 
reflect charged particles.  If these clouds converge, the 
particles gain energy over time.  If they diverge, the particles lose energy.
Later, it was realized \citep{Bell, Axford, krymskii, blandford}, that shock fronts are another 
site of Fermi Acceleration.  In the shock acceleration model, charged particles stream into 
magnetic perturbations in the post-shock region, reflect, and are scattered back across the shock by 
pre-shock Alfv\'en waves.  Repeated reflections steadily accelerate 
particles to a power law 
distribution.  This has been studied extensively (e.g. Jones \& Ellison 1991).
If the shock thickness is determined by the ion gyroradius then
ions are picked up out of the thermal population but electrons
must be injected at energies at or above the thermal energy by a factor of the ratio of proton 
to electron mass, $(m_p/m_e)$,
to incur power law acceleration.  The injection process is a critical outstanding problem for many applications 
of shock acceleration.  Shock Fermi acceleration is commonly referred to as first order
Fermi Acceleration because the sign of the energy gain after each cycle is positive and $dE/dt \propto v_c/v$, 
where $E$  and $v$ are the electron energy and speed, and $v_c$ the velocity of the magnetic compression.

Fermi acceleration can also take place in a fully turbulent plasma.  
There, the mirroring sites are turbulent perturbations, 
typically fast mode magnetohydrodynamic waves, randomly distributed throughout the plasma
(e.g. \cite{achterberg84}).  
Electrons encounter these perturbations such that
there is a stochastic distribution of energy gaining and energy 
losing reflections.  As is demonstrated below, there is a net  
dissipation of turbulent energy into high energy electrons.  Because the energy gaining and energy
losing reflections are equal to first order, STFA is often referred to as second order 
Fermi acceleration and proceeds more 
slowly than the first order process.  For STFA, $dE/dt \propto (v_c/v)^2$.

STFA has been considered extensively as the acceleration mechanism in impulsive solar flares, (e.g. \cite{LaRosa}).
Observations of these flares show hard X-ray emission with a downward breaking power law spectrum extending
from $10$keV to at least $0.5$MeV with the break energy narrowly distributed around $E_{br} = 100$keV and
a thermal distribution at energies below $10$keV \citep{Dulk, Krucker}.
The time structure of the emission shows distinct spikes of duration $~1$s and typical energy $10^{26}$erg
\citep{Aschwanden}.  Non-thermal emission occurs principally at the footpoints of the soft X-ray loop, and to a 
lesser extent at a loop-top hard X-ray source \citep{Tsuneta, Masuda}.  \citet{Brown} has shown that the emission
at a dense target is consistent with Bremsstrahlung radiation by electrons accelerated to a power 
law energy distribution
at some height above the target; in impulsive flares the acceleration 
site can be  associated with the loop-top region, while
the thick target is associated with the footpoints.   
We demonstrate that STFA is possibly responsible for the acceleration of electrons below $10$keV, while 
first order Fermi processes at the loop-top fast shock may produce the highest energy electrons.  In this  
picture STFA also provides power law distributed electrons in the range $10$keV$<E<100$keV, 
thus satisfying the shock injection criterion and producing the observed spectral break at $100$keV.  It is shown 
that in order to produce this spectrum the pitch angle scattering 
must obey the restriction that the scattering distance, $\lambda_P$, is 
inversely proportional to the energy of the scattered electron in 
the $10-100$keV range.  Matching the knee is difficult, requiring either
a cutoff in the secondary pitch scattering at $100$keV, or the sudden 
onset of yet another pitch scattering agent with a much shorter wavelength.  
While both are possible, these requirements provide a serious challenge to STFA models of electron acceleration.  

\begin{figure}
\epsscale{0.4}
\plotone{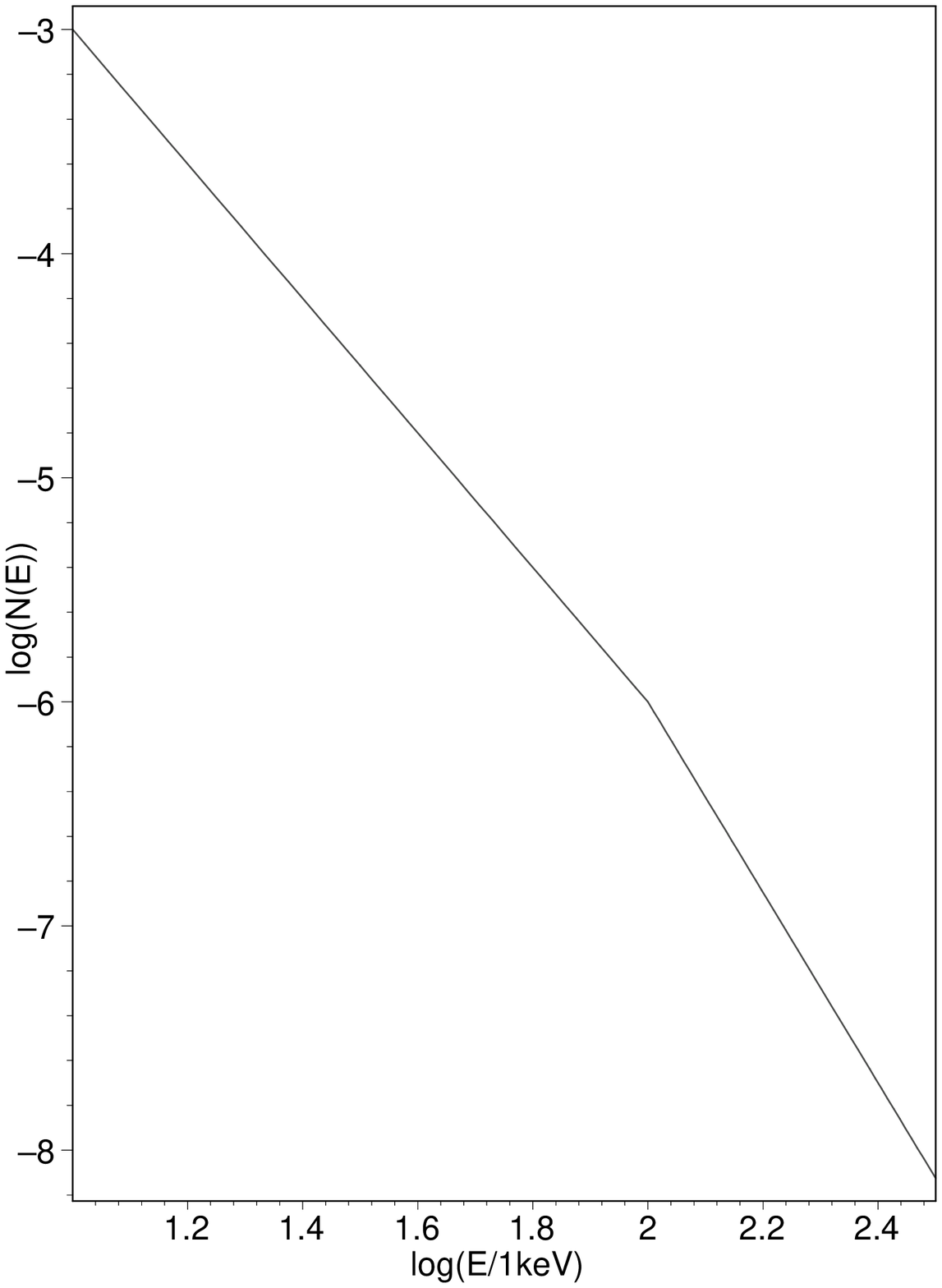}
\plotone{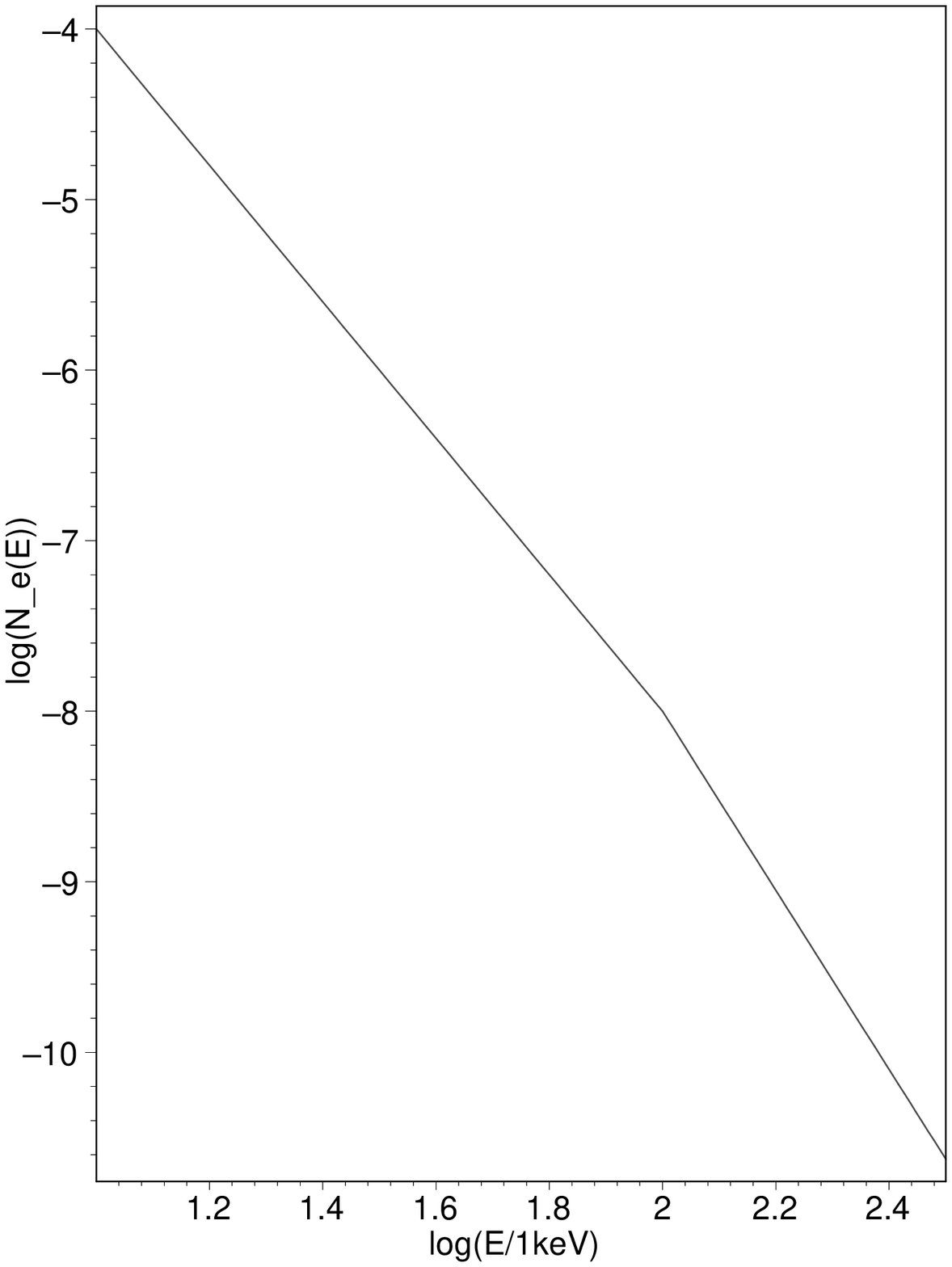}
\caption{The typical downward broken power law of an impulsive flare, as 
observed by \citet{Dulk}.  The left panel shows the hard X-ray spectrum, with $E_{br} = 100keV$, and spectral indices 
above and below $E_{br}$ of $4.25$ and $3$ respectively.  The right hand panel shows the electron spectrum in the emission region 
inferred from the given photon spectrum using a thick target Bremstrahlung model for the emission \citep{Brown, THE}.  
Again, $E_{br} = 100$kev and the spectral indices above and below $E_{br}$ are $4$ and $5.25$.}  
\label{spectrumsample}
\end{figure}

It is important to note that the downward break is not observed in all impulsive solar flares. 
Indeed, \citet{Dulk} observed a spread in break energies and spectral indexes.  Some of their flares exhibited no
discernible break, or even some upward breaks (ankles).
In this paper we address the fundamental process of 
STFA, and show how it can accommodate the downward breaking spectra observed in a subset of impulsive flares.  
The general form of the electron spectra we model is illustrated in figure \ref{spectrumsample}.  We assume a thick target
Bremmstrahlung emission model where the electron spectral index $\delta$ and the photon spectral index $\gamma$ are 
related by $\delta = \gamma +1$ \citep{Brown, THE}.  The data shown are mean values taken from the flares studied by \citet{Dulk}:
$E_{br} = 100$keV, $\delta = 4$ below $E_{br}$, and $\delta = 5.25$ above $E_br$. 

Furthermore, not all flares are observed to be dominated by electron acceleration.  A recent flare observed by \citet{Hurford}
 clearly shows regions of emission which are dominated by X-ray emission from electrons as well as regions which are 
dominated by gyrosynchrotron emission from MeV/nucleon ions.  Proton and ion emission appears to be associated primarily 
associated with larger flare loops.   \citet{MillerRoberts, Miller04} propose that this can be explained by a two stage 
process for ion acceleration.  First, ions are accelerated via gyroresonance to speeds of roughly $v_A$ by Alfv\'en waves, and 
subsequently are accelerated preferentially over electrons by magnetosonic waves.  They argue convincingly for the gyroresonant
acceleration by Alfv\'en waves on the basis of relative ion abundances.  However, it is unclear that the second stage
acceleration by magnetosonic modes must be resonant.  It appears that the second stage is consistent with STFA.  In any event, 
proton acceleration in long flare loops presents an interesting problem for acceleration models, in that suepr-Alfv\'enic protons
must be preferentially accelerated over electrons, but electron acceleration still must be dominant in shorter loops.  In this work, 
we presume that the loops are sufficiently short that protons remain sub-Alfv\'enic.  Longer loops, and the effects of proton 
acceleration on the shaping of the observed Bremsstrahlung X-ray spectra from high energy electrons, require further study. 

STFA is found to depend on two competing effects which we refer to as the steady and diffusive acceleration rates.  The
steady rate represents the net acceleration of electrons due to the slight advantage of head-on or energy gaining reflections
over catch-up or energy losing reflections.  The diffusive term represents the spreading of the electron distribution function
as a result of the stochastic nature of the reflections.
\citet{Longair} treats the two effects together using the Fokker-Planck equation.  Likewise, \citet{pandp} discuss general solutions of 
a simplified Fokker-Planck equation for STFA.  In a follow-up work, \citet{pps} apply their solution to solar flares.  While they 
produce spectra consistent with observations, they do not discuss the physics of electron escape.  
Furthermore, they focus mainly on the regime above $100$keV.   Some past treatments of STFA in impulsive flares 
focused exclusively on the diffusive term.  \citet{LaRosa} derives the diffusive acceleration rate using simple physical arguments.  
\citet{Chandran} derives the diffusive term using both phenomenological arguments and quasi-linear theory.  The latter also includes
Coloumb losses as a small correction; this is of note since the Coloumb loss term is mathematically similar to a negative steady 
acceleration term.  Herein we derive the steady term and compare it to the diffusive term,
finding the steady acceleration to be dominant in the non-relativistic
regime for impulsive flares.  As will be seen, our steady term differs
slightly from that of some previous calculations such as that in \citet{Longair}
in its dependence on the turbulent magnetic fluctuation strength.
This arises because we average only over the pitch angle phase space
for which STFA operates in magnetic mirroring, whereas
\citet{Longair} averaged over all pitch angles in considering a more
generic form of Fermi acceleration.  A similar averaging over all pitch angles is performed in \citet{Skilling, Webb}.

We first review the basic Fermi process using a test particle approach.  We then
derive an expression for the mean acceleration of electrons
in a turbulent plasma via STFA, and compare this to the diffusive STFA derived by \citet{LaRosa}.  Finally, we 
discuss the trapping of electrons in the turbulent accelerating region and show that at non-relativistic
energies, the electron spectrum depends strongly on the energy dependence of pitch angle scattering.  For scattering by
whistler wave turbulence, the emerging spectrum is quasi-thermal.  
In order to produce the $10-100$keV power law in solar flares, 
we show that there must be an additional source of pitch angle scattering which has a length scale inversely dependent 
on electron  energy; this mechanism is tightly constrained.  The existence of such a pitch angle scattering mechanism is 
yet to be determined.  This brings to the fore the most pressing difficulty with STFA models of electron acceleration; the pitch 
angle scattering requirements are stringent and might not be possible to meet.

\section {The Fermi Acceleration Process}

Consider a particle of charge $q$ traveling with gyroradius $r_G$ in a magnetic field 
of strength $B$ 
\citep{fermi49,Fermi,Spitzer}.
The charge follows a helical path, orbiting the field line while also moving parallel (or 
anti-parallel) to the field line.  Taking the condition for circular motion and the Lorentz force

\begin{equation}
F=q v_\perp B = \frac{m v_\perp ^2}{r_G}, 
\end{equation}
where $q$ is the electron charge,
and applying conservation laws for angular momentum and kinetic energy  
(${r_G} v_\perp$ and ${v^2}$ constant) yields

\begin{equation}
\frac{qL}{mE} = \frac{\sin^2{\theta}}{B}, 
\end{equation}
where $L$ is the angular momentum of the charge, $E$ the kinetic energy, and 
$v\sin{\theta} = v_\perp$ relates the total velocity to the component perpendicular to the 
field line.  The pitch angle, $\theta$, is the angle between the field line and the velocity vector.
As the charge enters a region of increasing $B$, such as a magnetic compression, the pitch angle 
evolves according to

\begin{equation}
\frac{\sin^2{\theta}_1}{B+\delta B} = \frac{\sin^2{\theta}}{B}
\end{equation}
where $\delta B$ is the increase in field strength and ${\theta_1}$ is the pitch angle at field
strength $B+\delta B$.  When $\sin{\theta_1} = 1$, the charge cannot penetrate further into the 
compression, and reflects.  This process, known as magnetic mirroring, is commonly used to confine
laboratory plasmas \citep{Dendy}.  It follows immediately that mirroring will not occur at a given compression 
unless the initial pitch angle satisfies 

\begin{equation}
\sin^2{\theta} \geq \frac {B}{B + \delta B}.  
\label{3}
\end{equation}
Fermi (1949, 1954) showed that moving magnetic mirrors, in particular molecular
clouds, can accelerate charges.  In the cloud's frame of reference (primed), mirroring
results in only a change in the sign of $v^{\prime}_\parallel$, the 
component of the initial velocity of the charge parallel to the field line in the compression's rest frame.  
Let us work for the moment in the limit where the compression speed and the particle speed are 
both $ \ll c$.
Transforming to the lab frame,  $\delta v_\parallel  = \pm 2v_c$, 
where $v_c$ is the drift velocity of the cloud.  The positive (negative) sign is for
head-on (catch-up) reflections between the charge and cloud.  Catch-up reflections are defined as those 
where the components of the compression and charge velocities parallel to the field line have the same sign.  
Head on reflections are those where the parallel components have opposite signs.  
The net change in energy from a reflection is given by 
\begin{equation}
\delta E_{\pm} = \frac{m}{2} (v^2_f - v^2_0) 
= \frac{m}{2}(2 \delta v_\parallel v_0 \cos(\theta) + (\delta v_\parallel)^2) 
	 = (2m)(\pm v_c v_\parallel + v^2_c),  
\label{Epm}
\end{equation}
where  $v_f$ and $v_0$ are the speeds after and before reflection, $m$ is the mass of the charge, and $v_\parallel = v_0 \cos(\theta)$.
Head-on reflections result in a positive energy change, while catch-up reflections can result in a 
negative energy change when $v_\parallel > v_c$.  In both types of reflection, there is the positive term 
proportional to $v^2_c$.   

One can repeat this derivation using Lorentz transformations in place of the Galilean transformations to 
generalize this result to particles of any energy scattered by compressions which are still restricted to 
non-relativistic velocities.  For a derivation see \citet{Longair}; we simply cite the result: 

\begin{equation}
\delta E_{\pm R} = 2 E \left[\pm \frac{v_c v_\parallel}{c^2} + \frac{v^2_c}{c^2}\right],
\label{Estep}
\end{equation}
where $c$ is the speed of light in vacuum.
Over time, charges trapped between converging magnetic compressions are subject to only head-on 
reflections, and are accelerated to higher energies.  

Notice that the change in momentum of a Fermi accelerated electron is solely in the component 
parallel to the mean field.  This corresponds to an increase in the electron's pitch angle.  As an 
extreme example, consider an electron of initial energy $E_0$ with pitch angle in the mean field $B$ 
approaching $\pi /2$.  Upon doubling the electron's energy via Fermi acceleration, the pitch angle in 
the mean field is reduced to $\pi/4$.  Clearly, acceleration to high energy must be accompanied by 
some additional scattering agent which isotropizes electron pitch angles on a short time scale, otherwise
Fermi acceleration shuts off after small accelerations as pitch angles evolve out of the range given by (\ref{3}).
The well known problem of pitch angle scattering remains largely unsolved
(e.g. \citet{achterberg1981, melrose, LaRosa}).  

Fermi acceleration is distinct from the transit time damping (TTD) treated, 
for example, in \citet{miller96}.  TTD 
is the magnetic analog of Landau Damping.  In TTD, electrons (or ions) 
which are near gyroresonance with waves of wavenumbwer $k$
are pushed towards the resonance by field gradients in the wave which alter the parallel component of the velocity.
Gradients in the electron velocity spectrum result in a net damping
or enhancement of the waves.  In the presence of a spectrum of waves, 
an electron can drift from resonance at $k$ to $k \pm \delta k$ and so 
forth, eventually reaching high energies.  Fermi Acceleration, however, 
is a non-resonant interaction.  Electrons will mirror at a compression 
regardless of energy provided that the pitch angle is sufficiently large.  
TTD is often referred to as resonant Fermi Acceleration because the 
two processes rely on similar physics.  Table \ref{tablescales} lists the relevent length and time scales 
for STFA in impulsive flares.

\begin{table}[t]
\centering
\begin{tabular}{cccc}                        
 \hline
Length scale        &  Description                                     & Time scale    & Description             \\ \tableline 
$L_T$               &  Turbulent outer scale                           & $\tau_S$      & Steady STFA acceleration time    \\
$\lambda_T$         &  Turbulent eddy scale                            & $\tau_D$      & Diffusive STFA acceleration time  \\
$\lambda_\parallel$ &  Parallel eddy scale                             & $\tau_{ED}$   & Turbulent eddy time         \\ 
$\lambda_\perp$     &  Perpendicular eddy scale                        & $\tau_p$      & Pitch angle scattering time \\
$\lambda_{SF}$      &  Effective STFA scale                            & & \\
$\lambda_p$         &  MFP for pitch angle scattering                  & & \\
$\lambda_{wh}$      &  $\lambda_p$ for scattering by whistlers         & & \\
$\lambda_{C}$       &  Constrained $\lambda_p$ for $10-100$keV         & & \\ \hline
\end{tabular}
\caption{Table of length and time scales relevant to STFA in impulsive flares.}
\label{tablescales}
\end{table}

\section{Stochastic Fermi Acceleration}

	We now consider the behavior of charges in a turbulent magnetic plasma where 
magnetosonic modes provide the sites of magnetic mirroring.  This 
scenario differs from first order Fermi acceleration by shocks in two ways.  
1) Consecutive mirroring events are not coherent, but rather stochastically distributed between 
head-on and catch-up.  
2)  The turbulent cascade governs the acceleration efficiency; the system 
picks out a scale where acceleration competes with the cascade.  In the solar corona plasma, 
$v_c$, the velocity of the magnetic compressions,  is the phase speed of the magnetosonic modes, 
which is roughly the Alfv\'en speed for the fast mode and the sound speed ($c_s$) for the slow mode.  
Typically, thermal electrons in the corona are 
super-Alfv\'enic and non-relativistic, $v_A \ll v_0  \ll c$, and $\beta \backsim 0.05$; 
we will solve the STFA problem in this regime.  

\subsection{Determination of the Steady Acceleration Rate}

Recall that the energy gain from a typical reflection is given in Eq.(\ref{Epm}) to be 
$\delta E_\pm = 2m(\pm v_\parallel v_c + v^2_c)$.  We define three parameters: $R$, the total rate of 
reflections; $R_+$, the rate of head-on reflections; and $R_-$, the rate of catch-up reflections.
The relation $R = R_+ +R_-$ is automatically satisfied by this definition as all reflections 
must be of either the head-on or catch-up type.  This allows us to write the approximate
acceleration rate as the sum of a coherent term and an incoherent term:

\begin{equation}
\left(\frac {dE}{dt}\right)_S = 2m[(R_+ - R_-) v_\parallel v_c + (R_+ + R_-)v_c^2],
\label{sdefine}
\end{equation}
where the subscript $S$ is used to distinguish our derived acceleration rate from that of \citet{LaRosa}.
The first term, proportional to $(R_+ - R_-)$ represents the mean acceleration due to the offset in 
the rates of the two types of reflection.  The second term, proportional to $(R_+ + R_-) = R$, 
the total reflection rate, represents the coherent term.  This expression
gives a full description of the mean acceleration of charges by the non-relativistic STFA process. 
To evaluate it in a particular plasma requires the determination of the head-on and catch-up
reflection rates $R_+$ and $R_-$.
  
	In order to obtain $R_+$ and $R_-$, consider the path a charge takes to encounter a
mirror.  Note that there is a distinction between an encounter and a mirroring because of the 
pitch angle condition for reflection.  We take the fraction of encounters which reflect to be $F$
and assume that this fraction is the same for both head-on and catch-up encounters: $F = F_+ = F_-$.  
This assumption is often taken in the regime where $v \gg v_c$ 
\citep{LaRosa}.  While this assumption is not strictly true, the effects of relaxing it are negligible.  In
Appendix B we repeat this calculation without assuming $F_+ = F_-$.
There is a well defined mean distance between encounters, 
$\lambda_T$ as well as a relative velocity between the particle and the compression 
$v_{\pm} = v_\parallel \pm v_c$.   The small difference in the head-on and catch-up speeds is 
responsible for the offset in rates.  For each type of encounter, the mean separation is $2 \lambda_T$ and the 
rate of reflections of any sort is specified by the general relation $R = F v_{rel} \lambda$, with $v_{rel}$ the 
relative velocity between the particle and compression, yielding

\begin{eqnarray} 
\nonumber
R_+ = F \frac{v_\parallel + v_c}{2\lambda_T} \\ \label{erates} R_- = F \frac{v_\parallel - v_c}{2\lambda_T} \\ \nonumber
 R = R_+ + R_- = F \frac{v_\parallel}{\lambda_T}. 
\end{eqnarray}
The offset in rates is thus 

\begin{equation}
R_+ - R_- = F\frac{v_c}{\lambda_T},
\end{equation}
and we can rewrite the average acceleration rate as

\begin{equation}
\left(\frac {dE}{dt}\right)_S 
= 2mF\frac{v_\parallel v^2_c}{\lambda_T} 
+ 2mF\frac{v_\parallel v^2_c}{\lambda_T} = 4mF\frac{v_\parallel v^2_c}{\lambda_T},
\label{ratecalcfirst}
\end{equation}
with the associated acceleration time scale $\tau_S = E/\left(\frac {dE}{dt}\right)_S $.
We thus see that the acceleration due to the offset in head-on and catch-up reflection rates and 
the acceleration from the coherent term in $v_c$ are equal.  

To simplify the derivation of the electron spectrum, we recast (\ref{ratecalcfirst}) as
\begin{equation}
\left(\frac {dE}{dt}\right)_S = 4mv^2_c R = \frac{dE}{dM} \frac{dM}{dt},
\label{11}
\end{equation} 
where we used (\ref{erates}) and define $M$ as the total number of reflections experienced by an 
electron, $dM/dt = R$, and 
\begin{equation}
dE/dM = 4mv^2_c.
\label{dedm}
\end{equation}

The quantity $dE/dM$, the mean acceleration of an electron per reflection, 
will be of particular use when examining electron escape in section \ref{sec:spec}.

\subsection{Comparison to the Diffusive Acceleration Rate}

	A different approach was taken by \citet{LaRosa}.  They set $R_+ = R_-$ in the $v \gg v_c$ regime and 
studied the diffusion of particles through energy space via random walk.
The starting point in their calculation of the electron acceleration
was the timescale for the e-folding of a charged particle's energy in the turbulent plasma
\begin{equation}
\frac{1}{\tau_D} = \frac{1}{E} \left(\frac{dE}{dt}\right)_D = \frac{F}{N \delta t},
\end{equation}
where $N$ is the number of mirrorings required to double the particle's energy and $\delta t$ is the 
time between encounters.  They set $R_+ - R_- = 0$, and also dropped the
last term in $v^2_c$ in Eq \ref{Estep}.  From (\ref{sdefine}) and (\ref{ratecalcfirst}) 
it is clear that if one of their two assumptions is
valid the other must also apply, and $(dE/dt)_S = 0$ in that limit.  
Under these assumptions
$\delta E_+$ and $\delta E_-$ are equal in magnitude, and from the standard solution of 
an evenly weighted random walk $N = (E/\delta E)^2$.  The acceleration rate is then

\begin{equation}
\left(\frac {dE}{dt}\right)_D =  \frac{E}{\tau_D} = F\frac{\delta E^2}{E} \frac{v_\parallel}{\lambda_T}
		= \frac{8F}{\lambda_T} \frac{mv^2_c v^3_\parallel}{v_0^2},
\end{equation}
where $v_0$ is the total initial speed and the subscript $D$ is used to denote LaRosa et al's 
diffusive acceleration rate.

	To complete the calculation of the acceleration rates, we must obtain 
$F$ and average over pitch angles.  The minimum accessible pitch angle for reflection is related to $F$ by

\begin{equation}
F = \cos({\theta_{min}}) = {\left(\frac{\delta B}{B}\right)}^{1/2},
\label{Feq}
\end{equation}
where we have applied the reflection condition from (\ref{3}).  In the regime where the 
compression ratio $\delta B/B <<1$, which will apply to the plasma of interest, and taking the assumption that
pitch angles are isotropic gives
$\langle \cos({\theta})\rangle =  \cos({\theta_{min}})/2$ and $\langle \cos^3({\theta})\rangle =  \cos^3({\theta_{min}})/4$.  
We now write the averaged acceleration rates as

\begin{equation}
\left(\frac {dE}{dt}\right)_S = 4mF\frac{\left<v_\parallel\right> v^2_c}{\lambda_T} 
		 	      = \frac{2m}{\lambda_T} (v^2_c v) \left(\frac{\delta B}{B}\right),
\label{dedtnr}
\end{equation}

\begin{equation}
\left(\frac {dE}{dt}\right)_D = \frac{8F}{\lambda_T} \frac{mv^2_c \left<v^3_\parallel\right>}{v_0^2} 
		= \frac{2m}{\lambda_T} (v^2_c v) \left(\frac{\delta B}{B}\right)^{2}. 
\end{equation}

\begin{figure}
\epsscale{0.4}
\plotone{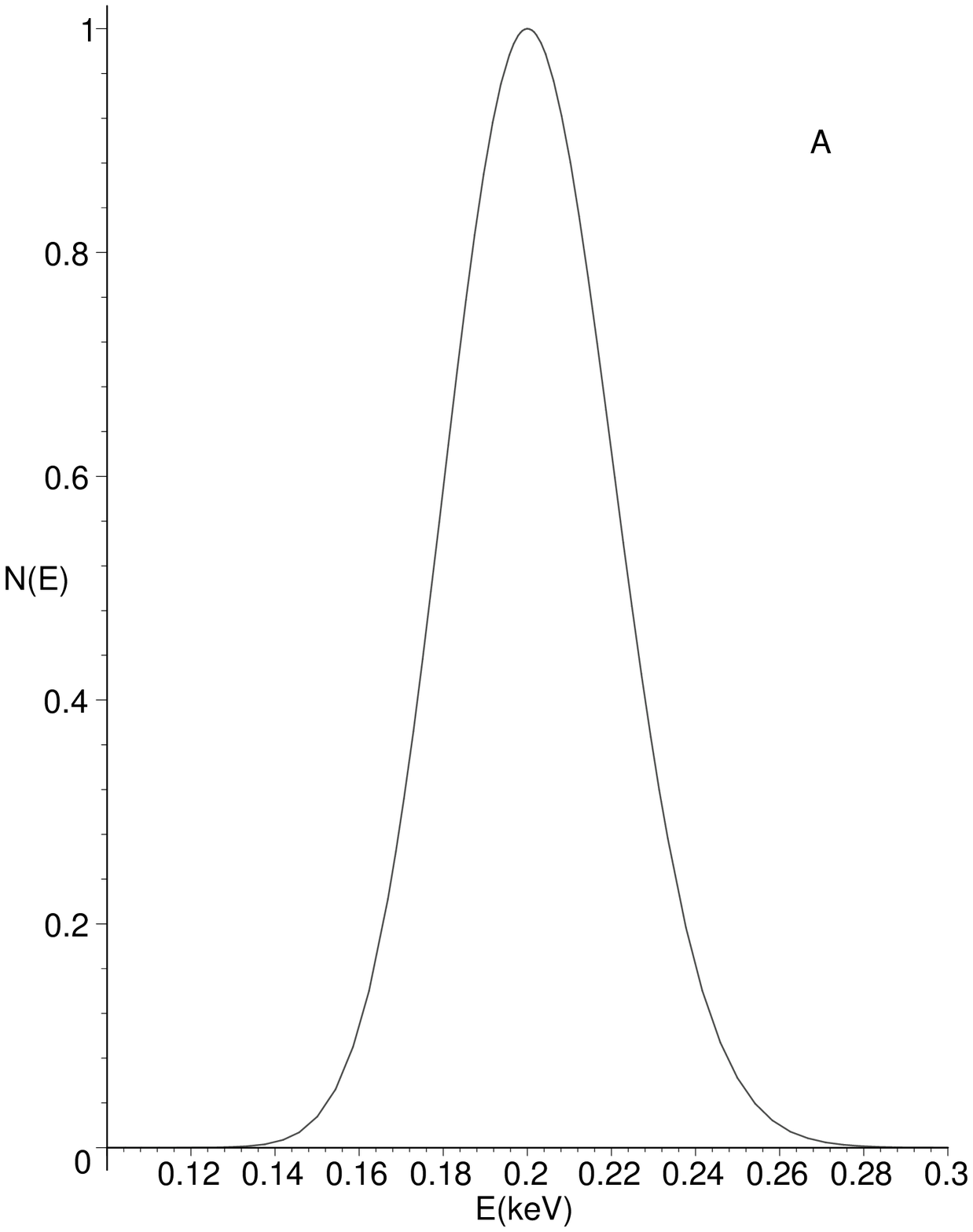}
\plotone{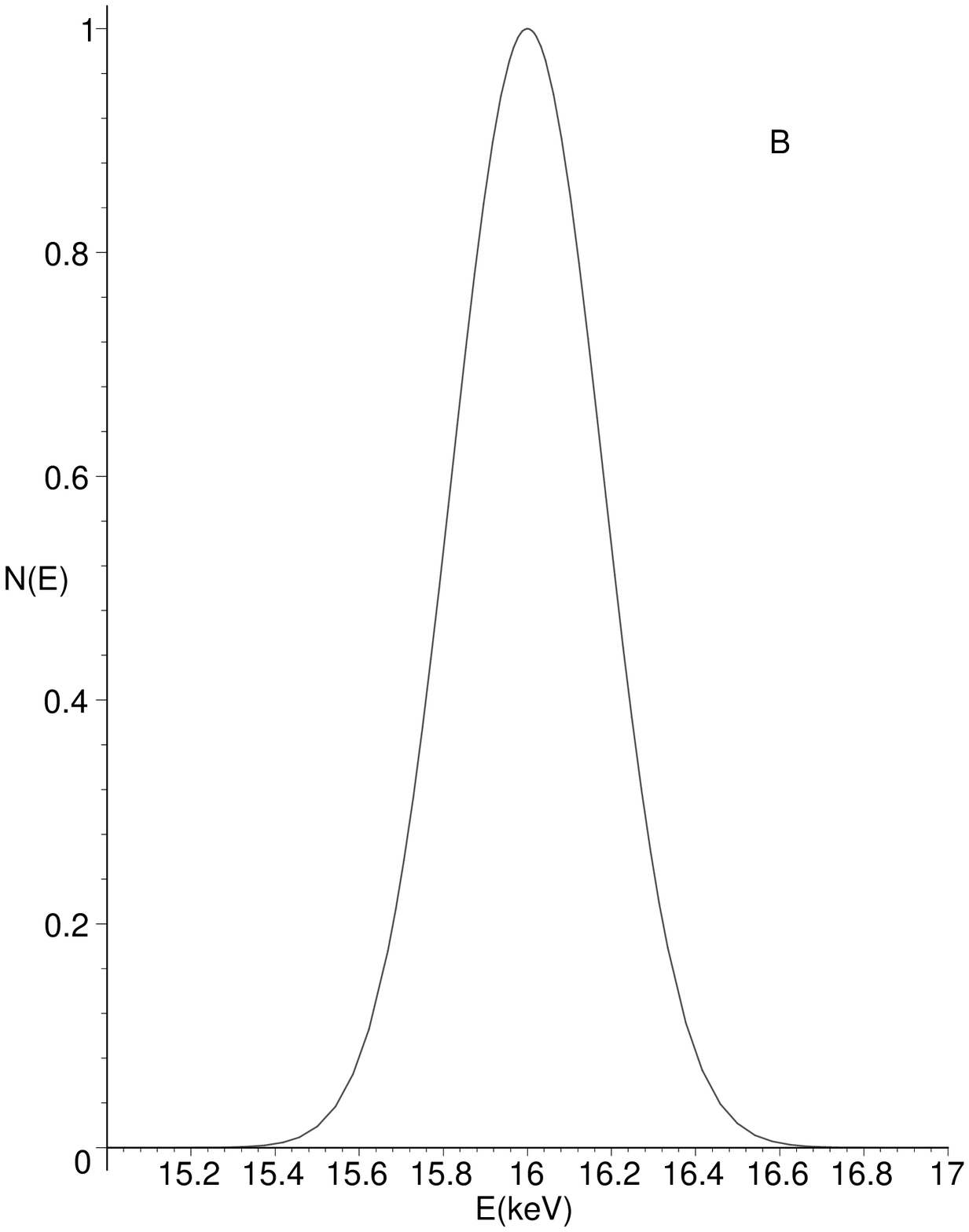}
\plotone{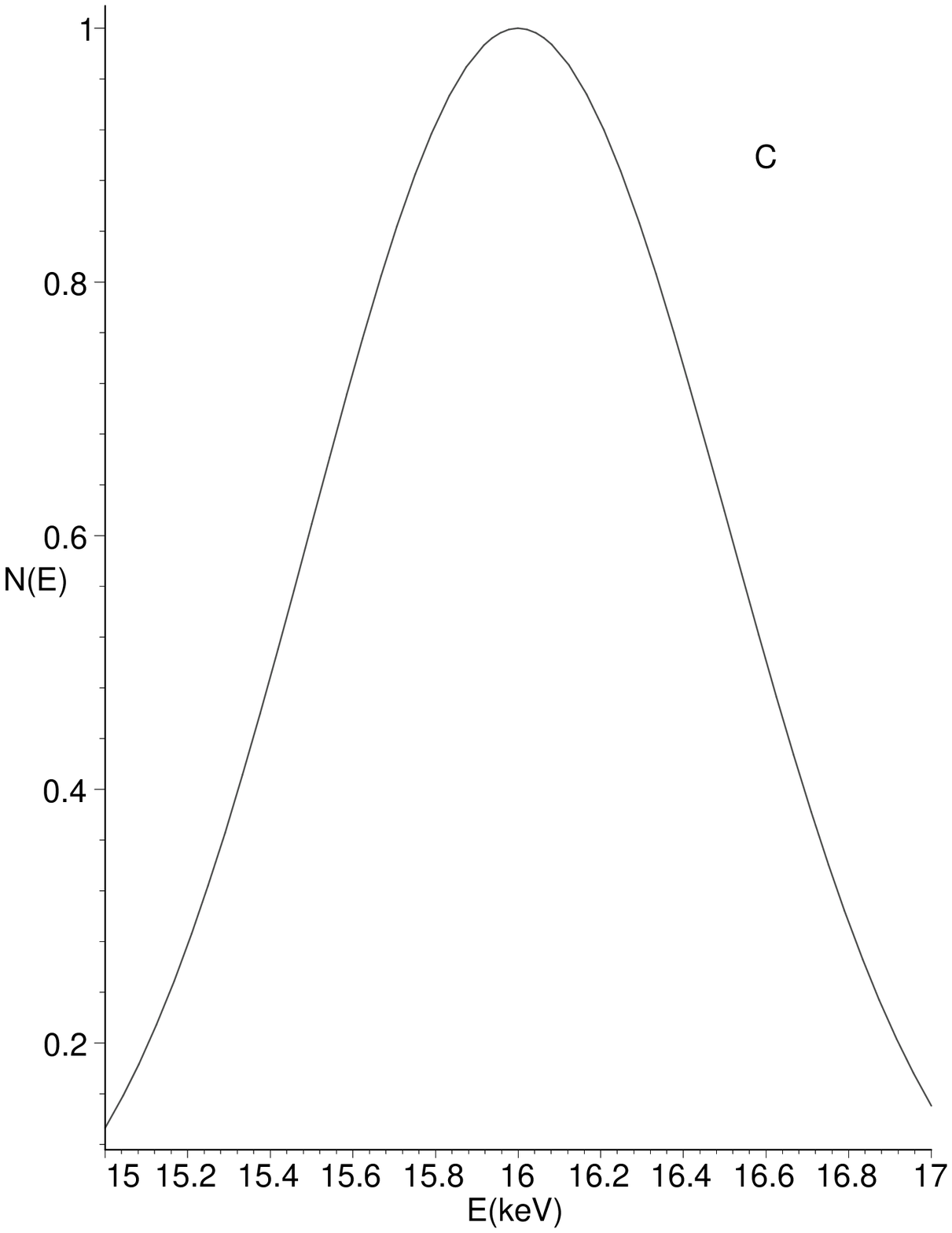}
\caption{The evolution of a sample electron 
energy distribution with an initially Gaussian velocity distribution.  The peak of the distribution is evolved from the
flare thermal energy, $0.2$keV to the post-STFA mean energy at $16$keV.  A)The  initial 
distribution function.  B) The same distribution after being evolved only by the steady process.  C) The 
distribution evolved through both the steady and diffusive processes.  Note that the relative width of the electron energy
distribution, $\Delta E/E_m$ dereases with increasing mean energy $E_m$.}
\label{evolved}
\end{figure}

	What do these two acceleration rates represent?  $(dE/dt)_S$ is the steady growth of the 
mean kinetic energy due to the drain of turbulence by the combined effects the slightly non-zero $(R_+ - R_-)$  
and the coherent $v^2_A$ 
term: a shift of the mean electron energy to higher energy.  On the other hand, $(dE/dt)_D$ represents the 
diffusion of energies away from the mean via random walk: a spreading of the distribution.  
The relative importance of the two is fixed by their ratio

\begin{equation}
\zeta = \frac{\left(\frac {dE}{dt}\right)_S}{\left(\frac {dE}{dt}\right)_D} = \frac{B}{\delta B}
\end{equation}
The combined result of the action of both processes on an initially narrow Gaussian energy
distribution is shown in Fig.\ref{evolved} where we have chosen $\zeta = 65$.
and assume that electrons do not escape.  Thus we can examine the 
evolution of electron energy spectra solely due to the influence of the two acceleration rates.  As $\zeta$ is 
increased, the steady (mean growth) term becomes increasingly dominant over the diffusive (distribution 
widening) term.  To understand STFA in a particular plasma,
both acceleration rates must be calculated.  In the event that the diffusive rate is very small 
compared to the steady growth rate, it can be ignored.  The steady growth rate is always faster than the diffusive 
growth rate for $a>1$.  As will be shown later, $\zeta \backsim 100$ in flare plasmas, and the diffusive term is negligible.

It is very important to note that our result differs from the standard for Fermi Acceleration, in which both the diffusive and steady
terms depend on the same power of $\delta_B/B$.  This difference arises as a result of the averaging over pitch angles.  To correctly
obtain $\left< \Delta E \right>$, one must only average over those encounters which result in a reflection.  For traditional STFA,the 
range of pitch angles which reflect
is ultimately determined by the turbulent magnetic field strength.  One factor of $\cos(\theta)$ in the expression to be 
averaged results in one factor of $(\delta B/B)^{1/2}$ in the acceleration rate.   
In other treatments, such as that of \citet{Longair, Webb, Skilling}, 
the acceleration mechanism is assumed to act at all pitch angles.  In this case, the averaging over $cos(\theta)$ while 
maintaining the asumption of pitch angle isotropy, yeilds a numerical value with no $\delta B$ dependence.  In this regime, the 
diffusive and steady acceleration terms have the same relative strength at all levels of turbulence.
 
\subsection{Specification of the Turbulent Cascade}

	This leaves $\delta B/B$ as the remaining parameter to be determined.
It is related to the turbulent length scale $\lambda_T$ through the cascade law.  Magnetohydrodynamic (MHD) turbulence proceeds
by the shredding of like sized eddies and subsequent formation of smaller eddies;  energy input into 
eddies on a large (outer) length scale,  $L_T$, cascades 
rapidly to smaller length scales on the eddy turnover time, $\tau_{ed}$, and finally dissipates 
at $\lambda_r$, the dissipation scale.  If the cascade obeys Komolgorov's steady state assumption then the 
energy flow through all length scales is constant, and independent of the scale.  This results in 
an inertial range between the scales $L_T$ and $\lambda_r$ where the turbulent energy density has a 
power law dependence on eddy size.  The draining of turbulence at eddy size $\lambda_r$ is usually determined
by a micro-physical process, such as resistivity.  When STFA is active, the turbulence can instead be drained by pumping energy into
electrons.  This sets another condition for STFA to proceed in a plasma: there must be some $\lambda_{SF}$ 
which is greater than the resistive length scale at which the STFA timescale is shorter than the cascade time, 
otherwise the turbulence will drain at the resistive scale before STFA can produce an appreciable electron
acceleration.  

There are three major MHD wavemodes: Alfv\'en waves, and the fast and slow magnetosonic waves.  Alfv\'en waves are purely 
transverse, and thus do not compress the magnetic field; they cannot participate in STFA.  Both the fast and slow modes are 
compressive, and are in principle capable of Fermi acceleration.  It has been argued that in low 
$\beta$ plasmas such as the solar corona, 
the slow mode is rapidly dissipated via Landau damping \citep{achterberg1981}.  However, more recent studies of MHD turbulence
indicate that the cascade time for GS turbulence is significantly shorter than the electron damping time, and 
slow mode damping by electrons can be ignored in turbulent flare plasmas \citep{Lithwick}.  A key difference in the two analyses is 
that \citet{Lithwick} treats MHD turbulence as inherently anisotropic, whereas \citet{achterberg1981} assumes isotropy. 
Furthermore, Maron (private communication) has shown in simulations which 
neglect damping that the slow mode may be driven with much higher total energy content than the fast mode
at low $\beta$.   A definitive resolution of the issue is beyond the scope of this paper.  However, we should point out that up
to this point, the calculation is independent of the choice of wave mode.    There is one significant difference between the 
two, however: slow modes propagate at roughly the sound speed ($v_c \backsim c_s$) while fast modes propagate at roughly the Alfv\'en
speed ($v_c \backsim v_A$).  It will be shown that due to the influence of pitch angle scattering, STFA is likely dominated by the fast mode.

MHD turbulence is in general anisotropic; 
the direction of any large scale mean magnetic field defines a 
preferred axis.  Also, even if the turbulence is isotropic
on large scales, smaller scales may see the larger scale turbulent
structures as an effective mean field.
\citet{GS} (hereafter, GS) modified the  
the Komolgorov assumption 
for the cascade of slow and Alvf\'en modes of MHD turbulence under the 
condition that the turbulence is anisotropic with 
scale $\lambda_\parallel$ along the field line and 
$\lambda_\perp$ perpendicular to the field line.  
The two directions are found to obey different 
cascade laws, with $\lambda_\parallel$ cascading more weakly than $\lambda_\perp$.  The parallel direction 
is of more importance to STFA, as it represents the  distance along the field 
line between 
reflection sites.  The GS power law for the parallel scale is \citep{GS, Lithwick}

\begin{equation}
\frac{\delta B}{B} = {\left(\frac{\lambda_T}{L_T}\right)}^{1/2},
\end{equation}
where $B$ is the mean magnetic field strength, and $\delta B$ is the turbulent field strength at 
parallel length scale $\lambda_T$.  

	The exact power law of MHD turbulence remains the subject of some debate,
so for now we assume a general power law of form
\begin{equation}
\frac{\delta B}{B} = \left(\frac{\lambda_T}{L_T}\right)^{1/a},
\end{equation}
where $a > 1$ is an arbitrary index.  Using the turbulent power spectrum and substituting for $F$ from 
(\ref{Feq}) we can rewrite the acceleration rates as

\begin{eqnarray} \left(\frac {dE}{dt}\right)_S = \frac{2m}{\lambda_T} v^2_A v 
{\left(\frac{\lambda_T}{L_T}\right)}^{1/a}
	= \frac{2}{L} m v^2_A v 
{\left(\frac{\lambda_T}{L_T}\right)}^{(1/a)-1} \label{tau} \\ \nonumber
{\rm\ so\ that \ } \  
\tau_S = \frac{1}{4} \frac{v L_T}{v^2_A} 
{\left(\frac{\lambda_T}{L_T}\right)}^{1- (1/a)} \\ \nonumber 
\left(\frac {dE}{dt}\right)_D = \frac{2m}{\lambda_T} v^2_A v {\left(\frac{\lambda_T}{L_T}\right)}^{2/a}
	= \frac{m}{L} v^2_A v {\left(\frac{\lambda_T}{L_T}\right)}^{(2/a) -1}\\ 
\nonumber
{\rm\ so\ that \ } \  
\tau_D = \frac{1}{4} \frac{v L_T}{v^2_A} 
{\left(\frac{\lambda_T}{L_T}\right)}^{1-(2/a)}.
\label{ratesarray}
\end{eqnarray}
For a typical turbulent cascade, where $\lambda_T <L_T$ and $a > 0$, $\tau_S < \tau_D$.

\subsection{$\lambda_{SF}$ and the Role of Pitch Angle Scattering}

	There remains the final step of determining the particular dissipation scale 
$\lambda_{SF}$ at which the energy drain takes place.  In order for STFA to overcome the cascade of
MHD turbulence, it must drain energy at a rate equal to the input rate at the outer scale. 
 If turbulent compressions are shredded and cascade faster than electrons can 
draw out energy via reflections, then the cascade continues to smaller length scales.  At  
smaller scales, STFA is more rapid.  STFA becomes competitive with the cascade at a scale determined
by $(dE/dt)_S = dE_T/dt$.  $(dE/dt)_S$ is acceleration rate of electrons and $dE_T/dt$ is 
the cascade rate of turbulent energy,  $ n m_p v^3_A/L_T$.  The STFA scale $\lambda_{SF}$ 
is different for acceleration by the two compressive MHD modes.  

For slow mode turbulence, $v_c = c_s$, and the balance is 

\begin{equation}
\frac{2nm_e}{L_T} c^2_s v {\left(\frac{\lambda_T}{L_T}\right)}^{(1/a) -1} = \frac{n m_p}{L_T} v^3_A. 
\end{equation}    
Solving for $\lambda_T/L_T$ and associating this particular $\lambda_T$ with $\lambda_{SF}$ gives

\begin{equation}
\frac{\lambda_{SF}}{L_T} = \left[\frac{1}{2} \frac{m_p}{m_e} \frac{v^3_A}{c^2_s v}\right]^{\frac{a}{1-a}}.
\end{equation}
In solar flares and a GS turbulent cascade ($a=2$), $v_0 = 1.2\times10^9$cm/s, 
$c_s = 3\times 10^7$cm/s and $v_A = 1\times10^8$cm/s \citep{LaRosa}, this gives $\lambda_{SF}/L_T \backsim 10^{-6}$. 
We have taken the initial electron velocity to be the mean velocity of the thermal background plasma.  The cascade will proceed down 
the inertial range to this length scale where STFA then acts as the micro-physical damping agent, 
rapidly draining the energy from turbulence into particles.   

In the case of the fast mode, where $v_c = v_A$, the rate balance is 

\begin{equation}
\frac{2nm_e}{L_T} v^2_A v {\left(\frac{\lambda_T}{L_T}\right)}^{(1/a) -1} = \frac{n m_p}{L_T} v^3_A, 
\end{equation}    
and the STFA length scale is then given by

\begin{equation}
\frac{\lambda_{SF}}{L_T} = \left[\frac{1}{2} \frac{m_p}{m_e} \frac{v_A}{v}\right]^{\frac{a}{1-a}}.
\end{equation}
For solar flare conditions, and a GS cascade ($a=2$), $\lambda_{SF}/L_T = 10^{-4}$.
We have tacitly assumed that the length scale for pitch angle 
isotropization is roughly equal to $\lambda_T$.  If it is not, the acceelration rate is retarded significantly, and STFA can 
be shut off.  To understand this we must further explore the role of pitch angle scattering.

As discussed above, pitch angle scattering is necessary during acceleration to maintain a population of 
electrons which satisfy the pitch
angle condition for reflection.  The strength of the pitch angle scattering strongly regulates the rate of acceleration.  
We consider three cases:
$\lambda_p \gg \lambda_{SF}$, $\lambda_p \ll \lambda_{SF}$, and $\lambda_p \backsim \lambda_{SF}$ where eddy 
$\lambda_p$ is the typical distance over which pitch angles are isotropized.   
In the first case, $\lambda_p \gg \lambda_{SF}$, electrons reflect a few times and quickly 
leave the pitch angle range in which they can reflect.  They then must stream a distance of order $\lambda_p$ 
before they can scatter again.  Thus the rate of 
reflections and the acceleration rate are both decreased by a factor of $\lambda_{SF} / \lambda_p$.  Since the acceleration rate
and cascade rate are not in balance, the cascade continues down to smaller scales $\lambda_T < \lambda_{SF}$.
 The nominal acceleration rate (eq \ref{ratesarray}) is proportional to $\lambda_T^{-1/2}$, while the 
retardation factor is proportional to $\lambda_T$.  The combined effect is a net acceleration rate which is proportional to
$\lambda_T^{1/2}$; smaller scale turbulence is actually less efficient as an accelerator.  As a result, STFA never turns on in this 
regime.     
In the second case, $\lambda_p \ll \lambda_{SF}$, the pitch angle scattering is far more rapid than acceleration.  
Since pitch scattering can take an electron
through $\mu = 0$, very strongly pitch scattered electrons traverse the plasma by random walking in steps of length $\lambda_p$, again 
reducing the rate of reflection, this time by a factor of $(\lambda_p / \lambda_{SF})^2$.  Unlike the previous case, this is not a 
problem 
for STFA; the retarding factor tends towards unity as the cascade continues to scales  $\lambda_T < \lambda_{SF}$.  
The net acceleration rate is proportional to $\lambda_T^{-5/2}$, and STFA turns on as the cascade proceeds to a sufficiently 
small scale.  In case three, where 
$\lambda_p \backsim \lambda_{S}$, pitch angle scattering and reflections proceed at the same rate.  Thus, electrons are capable of 
streaming freely from compression to compression, while they maintain a nearly isotropic pitch angle distribution.  This is the 
simplest pitch angle scattering regime for STFA.  

The identity of the accelerating wavemode is now easy to determine.  
In section \ref{sec:pesc} we show that whistler wave turbulence is a plausible
source of pitch angle scattering, at least for the lower energy quasi-thermal component of the spectrum.  At $3$keV, 
$\lambda_{wh}/L_T$ is roughly $10^{-4}$.  This places slow mode turbulence 
($\lambda_{SF}/L_T \backsim 10^{-6}$) well in the first regime.  Slow modes do not participate in STFA in these flares.  Fast modes, 
however, have $\lambda_{SF}/L_T \backsim 10^{-4}$ and therefore are in the nearly ideal range for acceleration.  Furthermore, both 
$\lambda_{SF}$ and $\lambda_p$ grow linearly with electron energy, so as electrons undergo STFA by fast modes in the presence 
of whistler wave turbulence, they remain in the same pitch angle scattering regime throughout.

\section{The post-acceleration spectrum}\label{sec:spec}

        The simplest case of STFA is the steady state, where we assume that electrons are injected into a
turbulent region at energy $E_0$ at a rate equal to that at which accelerated electrons escape.  The 
turbulent energy supply is continuously replenished at a large scale.  We are concerned with the energy
spectrum, $N(E)$, of electrons escaping the region.  Note that this is in general different from the 
spectrum of the electrons within the turbulent region.  We define $N_t(E)$ to be the 
total number of electrons reaching energy at least $E$ before escaping, such that 
\begin{equation}
N(E) \propto -dN_t(E)/dE.  
\label{N(E)fromN_t}
\end{equation}
Initially, we consider the case of strongly relativistic electrons;
a full derivation of this regime is presented in the Appendix.

To appreciate the calculational differences between the non-relativistic
regime of interest to solar flares and the more commonly studied
relativistic regime, we begin with the latter.
Following the approach used by \citet{Bell} for Shock Fermi acceleration, and writing $dN_t/dM = -p_{esc}N_t$,
where $p_{esc}$ is the mean probability of an electron escaping from the acceleration region, gives

\begin{equation}
\frac{dN_t}{dE} = -\frac{dN_t}{dM} \frac{dM}{dE} = -p_{esc}N_t \frac{dM}{dE} = -p_{esc}N_t \frac{1}{\alpha E},
\end{equation}
where on the right hand side we have for the moment taken the strongly relativistic limit: $dE/dM =\alpha E$ and 
assumed $p_{esc}$ to be constant.  This treatment of the highly relativistic limit follows \citet{fermi49}.    
One can solve for $N_t(E)$ by separating variables, integrating both sides and inverting 
the logarithms,  resulting in the familiar power law (see e.g. \citet{fermi49, Longair, jones1994})

\begin{equation}
N_t(E) =  N\left(\frac{E}{E_0}\right)^{- \frac{p_{esc}}{\alpha}},
\end{equation}
where $N$ is the total number of electrons.
From Eq. (\ref{N(E)fromN_t}), one obtains

\begin{equation}
N(E) = N_0 \left(\frac{E}{E_0}\right)^{-\left( 1+\frac{p_{esc}}{\alpha}\right)},
\end{equation}
where $N_0 dE$ is the number density of escaped electrons with $E = E_0$.
Notice that the logarithmic integrals in both $N_t$ and $E$ are vital to producing the power law.

For STFA by fast mode waves, the computation is more complicated because $p_{esc}$ is energy dependent. 
We solve for a general $p_{esc}$ in the non-relativistic regime, leaving the 
specification of the trapping for later discussion.  In the non-relativistic regime, the acceleration rate is not  
proportional to the kinetic energy as it is in the strongly relativistic regime.  Instead, one has 
$dE/dM = 4mv^2_A$.  We thus write, using (\ref{11}),

\begin{equation} 
\frac{dN_t}{dE} = -\frac{{N_t}{p_{esc}}}{4mv^2_A},
\end{equation}
which can be rewritten as

\begin{equation}
\frac{dN_t}{N_t} = \frac{p_{esc}}{4mv^2_A} {dE}.
\label{dndt}
\end{equation}
We have assumed that $dN_T/dM = p_{esc} N_t$.  This is reasonable as long as the electrons can be treated as statistically independent 
and collisionless.  In this case, it is reasonable to assume that $p_{esc}$ carries no inherent dependence on $N_t$ and the escape rate
is simly given by the product of the number of electrons in the volume and the mean escape probability.    
Taking $p_{esc} = p_0 (E/E_0)^{-1}$ allows us to solve for a particularly interesting $N(E)$.  We see immediately that

\begin{equation}
\frac{dN_t}{N_t} = -\frac{E_0}{E} \frac{p_0}{4mv^2_A} dE,
\end{equation}
and $N(E)$ is again a power law energy distribution:

\begin{equation}
N(E) = N_0 \left(\frac{E}{E_0}\right)^{-\left(1+\delta\right)},
\end{equation}
where $\delta = {p_0 E_0}/{4mv^2_A}$ .
However, in any other case, STFA does not produce a simple power law.  
The importance of the trapping mechanism is now clear; the combined energy dependence of the 
acceleration and escape must be $E^{-1}$ to produce a power law spectrum.  
Such a spectrum relies on the coincidental logarithmic integrals over both $N$ and $E$.  

\subsection{Calculation of $p_{esc}$}\label{sec:pesc}

Let us now calculate $p_{esc}$ for non-relativistic electrons within the turbulent volume, and consequently the energy 
spectrum of electrons.  For simplicity, 
let us take the turbulent region to be rectangular, with the long axis, $z$, parallel to the direction of the bulk flow, 
with $z=0$ and $z=L_F$ fixed to the
downstream and upstream boundaries of the region respectively.  $L_F$ is taken to be the extent of the region of turbulent flow, 
which is presumed to be the entire 
distance between the reconnection sheet and the top of the soft X-ray loop.  This distance is typically of size $10^{10}$ cm for 
solar flares \citep{Tsuneta}.  
The largest eddy size in the turbulence, $L_T$ is set by the width of the outflow, typically $10^8$cm.  
Thus the turbulent volume consists of a number of cells, each of which flows downward from 
the reconnection point towards the loop-top. An electron escapes the acceleration region only when it reaches the X-ray 
loop at the base of the turbulent region.  These individual cells may be associaetd with single bursts or fragments of X-ray
emission, and thus are responsible for the temporal structure of impulsive flares.
In order to escape the region with energy $E(M)$, an electron must stream from its location 
in the region at some height $z$ to the boundary at $z=0$ after the $M$th reflection
without further reflection.  We will assume that the electrons are contained in the region in the $x-y$
plane by gyration around large scale field lines.  To further simplify the problem, we shall assume that the 
electron density remains uniform throughout the turbulent region.  We also neglect the bulk flow speed, 
$v_f = 8 \times 10^7$ cm s$^{-1}$ \citep{Tsuneta} since the legth of the downflow region is roughly $10^{10}$cm.  This gives a
flow time from the reconnection region to the loop-top of $100$s.  The acceleration process is fixed to the much shorter   
$1$s time scale by the temporal size of the observed energy release fragments and the MHD eddy turnover time.  Thus, bulk flow into 
the flare loop is not likely to be a dominant process in cutting off the acceleration. 

Take the mean $z$-component of the distance streamed between reflections to be $\lambda_z$; $\lambda_z$ carries an energy 
dependence inherited from the energy dependence of the pitch scattering.   The probability of 
escaping at $z=0$ after the $M$th reflection from a point at height $z$ is given by 

\begin{equation}
p_{esc}(z) = \frac{1}{2} e^{-z/\lambda_z},
\end{equation}
and the mean escape probability of electrons distributed uniformly across the length of the region is

\begin{equation}
p_{esc} = \frac{1}{L_F} \int_0^{L_F} p_{esc}(z) d{z} = \frac{\lambda_p}{L_F} (1-e^{-2L_F/\lambda_p}),
\label{pesc}   
\end{equation}
where we have taken $\lambda_z = \lambda_p/2$
from the isotropy in pitch angles, with $\lambda_p$ the pitch angle scattering length scale.  To obtain the 
spectrum of solar flare electrons requires specification of the pitch angle scattering.

Both \citet{miller96} and \citet{LaRosa} assume strong scattering, and 
suggest that the scattering agent above $1$keV is resonant interaction with lower hybrid (LH) turbulence, or
circularly polarized electromagnetic waves, such as whistler waves.  Below
1keV, Coulomb interactions are thought to be sufficiently strong to isotropize the electrons.  It should be noted that 
\citet{melrose} sets the threshold for the whistler mode resonance at $25$ keV for flare plasmas, while \citet{millerstein} argue 
that the resonances 
extend down to $1$keV.  We assume the latter.
In a recent study, \citet{Luo} considered whether LH wave turbulence is the primary mode of electron acceleration in solar flares.
They concluded that the pitch angle scattering is too inefficient to maintain isotropy. Thus, we assume that LH wave turbulence 
cannot supply sufficient pitch angle scattering to sustain STFA either. Whistler waves are more promising.

\citet{melrose} associates the frequency of pitch angle scattering $\nu$ with the pitch angle diffusion coefficient in the quasi-linear 
equation.  Thus, $1/\nu$ is the chracteristic time scale for effective 
pitch angle isotropization of the electron distribution.  It should 
be noted that, in general, some small anisotropy is likely to remain in the distribution, and that this anisotropy could be responsible 
for the generation of the whistler waves.  However, the source of these waves is still uncertain.   From \citet{melrose}, we have that

\begin{equation}
\nu = \frac{\omega^2_p \epsilon(\omega_R)}{\gamma_e \Omega_e n_e mc^2},
\label{nu}
\end{equation}
where $\gamma_e \backsim 1$ is the Lorentz factor of the electron, $\Omega_e$ is the electron gyrofrequency, $\omega_p$ is the 
plasma oscillation frequency, and 

\begin{equation}
\epsilon(\omega_R) = \frac{m_p}{2} n_p v^2_A \frac{\delta B}{B},
\end{equation}
is the energy density of the turbulence at the resonant wavelength. This allows us to rewrite (\ref{nu}) as  

\begin{equation}
\nu = 5 \times 10^{7} B_{100} n^{-1/2}_{10} \left(\frac{\lambda_T}{L_T}\right)^{1/2},
\end{equation}
where we have used $v^2_A = B^2/4 \pi n_p$, $\Omega_e = 1.8 \times 10^9 B_{100}$, $\omega_p = 5.7 \times 10^9 n^{1/2}_{10}$, 
and the dimensionless parameters $B_{100} = B/100$G and $n_{10} = n_e/10^{10}$cm$^{-3}$.
 
We must compare $\nu$ to the growth time for pitch angle anisotropy due to STFA. 
Bearing in mind that for STFA, $dv_{\perp}/dt = 0$, 

\begin{equation}
\frac{dE}{dt} = mv_{\parallel} \frac{dv_{\parallel}}{dt},
\end{equation}
and the pitch angle evolves according to

\begin{equation}
\frac{d(\cos{\theta})}{dt} = \frac{1}{m} \frac{sin^2{\theta}}{v^2 \cos{\theta}} \frac{dE}{dt}.
\end {equation}
Substituting in from equation (21)for $dE/dt$ and assuming a GS cascade ($a=2$), gives 

\begin{equation}
\nu_{SF} = \frac{d(\cos{\theta})}{dt} = \frac{2}{L_T}\left(\frac{\lambda_T}{L_T}\right)^{(-1/2)} \frac{v^2_A}{v} \frac{1}{\cos{\theta}}.
\end{equation}
For impulsive flares, $E_0 = 0.3keV$, $B_{100} = 2$, and 
$n_{10} = 1$.  Taking $\lambda_T = \lambda_{SF}$ results in $\nu = 1 \times 10^{6}$ s$^{-1}$ and 

\begin{equation}
\nu_{SF} = 2 \times 10^2 \left(\frac{E_0}{E}\right)^{1/2} s^{-1}, 
\end{equation}
where $\nu_{SF}$ is evaluated at the threshhold pitch angle for reflection.  To maintain pitch angle isotropy, scattering by whistler 
waves must occur on a time scale shorter than pitch angle evolution by STFA.    
Thus, as long as $\nu_{SF} < \nu$, isotropy can be maintained.  
This condition is met for all $E > E_0$.  Whistler modes, if present, are capable of providing sufficient pitch angle 
scattering to maintain isotropy.  

In addition to maintaining pitch angle isotropy, the scattering mechanism must also operate at a length scale which traps electrons
in the volume; $\lambda_{wh} \ll L_T = 10^8$cm must be satisfied, or else electrons rapidly leave the acceleration region and STFA 
shuts off.  We obtain the pitch scattering length scale for whistler waves, $\lambda_{wh}$,

\begin{equation}
\lambda_{wh} = \frac{v}{\nu} = 2 \times 10^3 \left(\frac{E}{E_0}\right)^{1/2} cm. 
\end{equation}
 The required condition is satisfied for energies below $100$kev.

\subsection{The Electron Spectrum and Constraints on the Secondary Pitch Angle Scattering}

\begin{figure}
\plotone{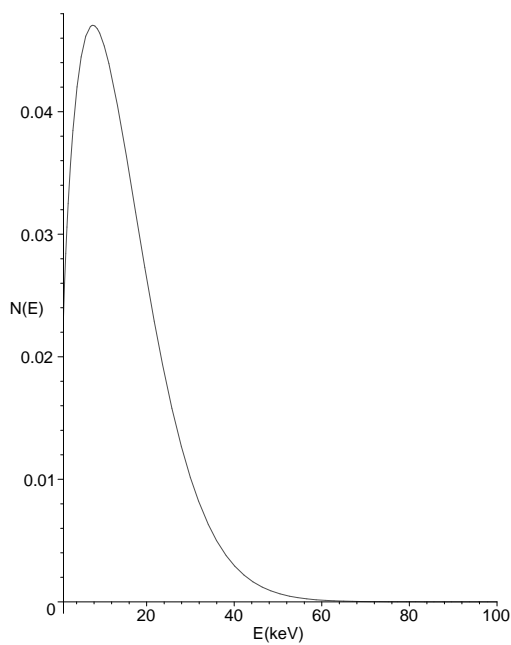}
\caption{The spectrum of non-relativistic STFA under impulsive flare conditions with whistler wave turbulence as the only pitch 
angle scatterer. $E_0 = 0.3$keV.}
\label{LHspec}
\end{figure}

In order to obtain the spectrum of the escaped electrons, we can now substitute the functional form for $p_{esc}$ from 
(\ref{pesc}) into (\ref{dndt})

\begin{equation}
\frac{dN_t}{N_t} = \frac{p_{esc}}{4mv^2_A} {dE} = \frac{\lambda_p}{L_F} (1-e^{-L_F/\lambda_p})\frac{1}{4mv^2_A} {dE}.
\end{equation}
Next, by choosing $\lambda_p = \lambda_{wh}$, we use $\lambda_p/L_F = A E^{1/2}$, with $E$ in units of keV, and rearranging, obtain

\begin{equation}
N(E) = \frac{dN_t}{dE} =  AE^{1/2} (1-e^{-1/AE^{1/2}})\frac{1}{4mv^2_A} {N_t}.
\end{equation}
The resulting  spectrum, $N(E)$, is plotted in figure \ref{LHspec}.
This is consistent with the thermal component to the flare spectrum observed using RHESSI \citep{Krucker}.  

In addition to the thermal component, RHESSI observations show a clear power law region extending from roughly $10$keV up to 
at least $50$keV, above which the data are uncertain, but consistent with a continuing power law.  Previous observations using the 
ISEE3/ICE instrument also show a power law throughout the range of the instrument, $25-300$keV; the spectrum typically breaks downward 
at $100$keV \citep{Dulk}.  More recent observations with RHESSI could push the low energy threshold for the power law as high as $35$keV \citep{Holman}.
The spectral index below the break is $\backsim 3$, while above the break it is $\backsim 4$.
Elsewhere (Blackman (1997); Selkowitz and Blackman (2004) in preparation), we discuss first order acceleration at the loop-top 
fast shock.  Fast shocks are well known to accelerate super-thermal 
particles to power law energy spectra, even in the non-relativistic limit \citep{Bell}.  However, in order to be accelerated, 
electrons must satisfy the requirement that $E \gg (m_p/m_e) v^2_{s} = 10$keV in solar flare plasmas, where $v_s = 10^8$cm s$^{-1}$ is
the inflow speed of the plasma at the shock \citep{Blackman}.  This places the injection energy at rougly $100$keV.  
The correspondence of the shock injection energy and the observed break energy is noteworthy.  
A possible mechanism to reproduce the observations is for STFA to produce a power law spectrum in the $10-100$keV regime
which then satisfies the injection criterion for loop-top fast shocks.  
Since STFA by magnetosonic turbulence in the presence of whistler wave turbulence pitch scattering is insufficient to 
produce the power law component.  However, it is possible that a second pitch scattering agent exists which produces the power law 
in the $10-100$keV range.  We examine the constraints imposed on this scattering.  

To produce a power law spectrum, non-relativistic STFA requires  $p_{esc} \propto E^{-1}$.  From (\ref{pesc}), we see that this is 
true only if $\lambda_p/L_F = \Gamma/E$, where $ 2E/\Gamma \gg 1$
and the  exponential term is small.  Here $\Gamma$ is a constant parameter which fixes the strength of the pitch scattering.
While the physics of the pitch angle scattering is not well understood, this constrains the 
scattering mechanisms available to STFA.  We define $\lambda_C = L_T\Gamma/E = 2\times 10^7 (E_0/E)$cm to be the pitch scattering 
length scale of the constrained mechanism, and the electron spectum is given by

\begin{equation}
N(E) = N_0 \left(\frac{E}{E_0}\right)^{-\left(1+\Gamma/(4mv^2_A)\right)}.
\end{equation} 
$\Gamma$ is constrained by the observed X-ray spectral index of $\gamma = 3$.  It is a standard prediction
of flare models \citep{Brown, Tsap, Kiplinger} that the spectrum of electrons accelerated
above the loop-top is steeper than the spectrum of the thick target Bremstrahlung X-rays emitted at the footpoints in solar 
flares.  We assume that to match the RHESSI X-ray data requires an index  $\backsim 4$ for the electrons, or 

\begin{equation}
\Gamma = 12mv^2_A = 0.072keV = 0.17E_0.
\end{equation}
The exponential term in $p_{esc}$ is indeed small as $2E/\Gamma = 270$ at $E=10$keV.   The electron spectral index could conceivably be 
as high as $6$, in which case $\Gamma = 0.28E_0$, and the exponential can still be safely neglected. 
 
It is insufficient to merely produce the proper power law.  The scattering agent must also be able to
 reproduce the transition from thermal to power law
spectrum at the correct energy, $E_c$.  We recall $\lambda_{wh}$ to be the length scale of pitch angle scattering associated with 
whistler wave turbulence.  In impulsive flares, $\lambda_p = \lambda_{wh}$ below $E_c$.  Above $E_c$, $ \lambda_p = \lambda_C$. 
To obtain $E_c$ one sets $\lambda_{wh} = \lambda_C$.  $E_c = 23$keV, which is 
consistent with the observed threshold of $\backsim 10$keV.

There is one more important constraint imposed by the observations: the knee at $100$keV.  \citet{Dulk} demonstrate a distinct downward 
break in the power law spectrum at roughly $E_{br}=100$keV.  The break energy varies somewhat from flare to flare, but is consistently 
observed in all of the impulsive flares in their sample.
 Unlike downward breaks, upward breaks are easily 
explained by the meshing of two acceleration mechanisms, as the shallow component which dominates above the break emerges naturally 
from beneath the steeper power law which dominates below.  For upward breaks, $E_{br}$ is the naturally occurring crossover point.  
The matching problem is much more difficult for knees in the  absence of significant cooling on timescales of interest.  Since the 
steep component is above the break energy and the shallow component below it, both must be truncated at the break energy.  If either 
one extended beyond the break, then that one would overrun the other, and there would be no break at all.    
The most natural solution for a knee is a single acceleration mechanism which undergoes some transition at the break energy.  One 
such example is the knee found by \citet{Bell2} in the spectra of shock accelerated electrons at roughly $1$GeV.  
This knee results from the transition from the non-relativistic to 
the relativistic regime.  There is no apparent natural transition for STFA of electrons at $100$keV.
However, there is a well defined low energy cutoff for a power law spectrum at $100$keV, the shock injection energy.  The shock 
injection threshold is not only at the right energy, but is also a variable cutoff, depending on the ion temperature 
and local magnetic field strength, consistent with the variability in the observed $E_{br}$.

\citet{Bell2} has shown that shock acceleration does not change the spectrum of electrons if the
pre-shock spectrum is shallower than the post-shock spectrum which obtains from a steep pre-shock spectrum.
Shock Fermi acceleration cannot steepen a power law spectrum; it can only 
make it shallower.  This is another difficulty for knee matching.  The STFA spectrum must have a sharp cutoff at $E_{br}$ in order to 
match the knee.  This may not be an 
impossible condition to meet, especially as $E_{br}$ may be greater that the injection energy, not precisely equal to it.  

One natural cutoff occurs when $\lambda_C = \lambda_r$; acceleration will shut off when the cascade reaches the resistive scale.  
For slow modes in impulsive flares, $\lambda_r = 10^3$cm 
\citep{LaRosa, Chandran, Lithwick}, which for $\Gamma = 0.073$keV gives a cutoff energy of $7\times 10^5$ keV, which is both too high 
and far outside of the non-relativistic regime.  
A possible solution is that the constrained pitch scattering has a maximum 
resonant threshold at $E_{br}$.  Another possible solution is that yet another very strong pitch scattering 
mechanism has a threshold energy of $E_{br}$ and a length scale $<\lambda_r$.  Both of these solutions 
are presently ad hoc.  This underscores the need for a more thorough understanding of pitch angle scattering in astrophysical plasmas.
It also illustrates the limitations of STFA as an acceleration mechanism in solar flares;  if STFA alone were to account for the 
spectrum from $10-100$ keV, the tight constraints on the pitch angle scattering mechanism that we have identified are required.

\section{Summary and Discussion}

STFA in the non-relativistic limit behaves differently from highly relativistic STFA.  At the core of these differences is the energy 
dependence of the electron velocity at low energies.  Thus, unlike the relativistic case, both the rate of reflections 
and the probability of escaping the acceleration region at an energy $E$ vary.  Using a test particle approach, we have examined this 
behavior and derived the spectrum of post-acceleration electrons in a plasma under impulsive solar flare conditions. 

For traditional STFA, where there is a minimum pitch angle constraint which determines whether an individual encounter results in
reflection, it is seen that the steady acceleration rate can dominate over the diffusive acceleration rate.  This arises from the 
averaging over pitch angles to evaluate $\left< \Delta E \right>$.  Some previous treatments of the generalized Fermi acceleration 
problem do not have such reflection conditions, and thus do not retain factors of the turbulent field strength, $\delta B/B$, when
averaging.  In those treatments, such as \citet{Longair, Skilling, Webb}, the steady and 
diffusive terms typically are seen to be of the same order.
For some processes this is appropriate, however non-resonant STFA is not one of them.
Thus, the phase space conditions for scattering by the acceleration mechanism can play a very significant role, even in cases where 
pitch angle isotropy is maintained.  

The nature of the pitch angle scattering turns out to be the dominant factor in determining electron escape, and therefore the shape of 
the spectrum. 
We find that whistler wave turbulence, which is well studied in solar flares \citep{melrose, millerstein}, is an excellent 
source of pitch angle scattering which allows STFA to produce a quasi-thermal electron distribution that peaks at $E \approx 5$keV.  
This matches the lowest energy portion
of the observed X-ray emission very well.  
However, to produce the power law spectrum observed in the range $\backsim 10-100$keV by STFA 
requires at least a second scattering
mechanism.  Matching the spectral index and the transition energy from quasi-thermal to power law spectrum requires an undetermined 
scattering mechanism which satisfies $\lambda_C / L_F = \Gamma/E$ with $\Gamma = 0.073keV$, and naturally becomes the dominant pitch 
angle scatterer at roughly $20$keV.  

If the constrained pitch angle scattering mechanism is discovered, it implies that the acceleration of electrons in solar flares 
is at least a two stage process.  The first stage, 
STFA in the downflow region, produces both the quasi-thermal spectrum below $\backsim 10keV$ and the lower half of the power law 
spectrum up to $100$keV.  To produce the 
highest energy electrons, as well as the spectral break at $E=100$keV requires a second 
acceleration mechanism at the top of the soft X-ray loop.  We are further exploring the possibility that 
first order acceleration at a weak fast shock, formed as the downflow impacts the top of the closed flare loop, is responsible for 
electron acceleration to the highest observed energies.  
Acceleration at fast shocks is known to have an injection energy of roughly $100$keV, and varies with temperature.  This coincides 
with the break in the power law spectrum at $100$keV, and is consistent with the variability observed by \citet{Dulk} in $E_{br}$.

Recently, Chandran (2003) concluded, using quasi-linear theory,
that STFA for slow modes is not viable in the 10-100keV regime.
While we also find slow modes to be ineffective, differences between our paper and 
\citet{Chandran} must be kept in mind.    
Chandran (2003) assumed that
$dp/dt \propto  p$ for STFA. While this is true in the strongly
relativistic limit for STFA, we do not assume that this
is true in the lower energy regime (see eq. 22).
Second, unlike Chandran (2003) we do not assume herein that $P_{esc}$ has
to be energy independent.  These two assumptions play a significant role in shaping electron spectra.

Another concern which can be raised about the effectiveness of STFA as the electron acceleration engine 
in impulsive flares is the total energetics of the process.  Since STFA, as developed above, only is efficient in a short length 
scale regime where one also has $\delta B / B \ll 1$ it might seem that only a 
small fraction of the released flare energy is available for electron acceleration.  This is not the case.  The total energy 
contained in single turbulent cell is given by $(1/2)m_p v^2_A n L^3 \backsim 10^{26}$erg, where $n = 10^{10}$cm$^{-3}$ is the electron 
number density in the flare plasma.  The energy in a 
single turbulent cell is similar to the energy contained in one X-ray emission fragment.  Although only a fraction of the energy 
in one turbulent cell is ever at $\lambda_p$ at one time, it does all cascade down to $\lambda_p$ over an eddy turnover time.  
Thus, while $\delta B / B$ is  always small, the energy throughput can still be high enough to accelerate the  
electrons.  The similarity in total energy between a single turbulent cell and an individual impulsive X-ray fragment strongly suggests 
that the two are related.  

\citet{Miller97} estimates that as much as $\simeq 94 \% $ of the magnetic energy in a flare is available in the turbulence, 
which is sufficient to produce the high energy electrons inferred from the observed X-rays, but raises concerns about the efficiency of 
STFA, particularly in competition with other sources of dissipation.   
While we have not fully 
studied other dissipation mechanisms which might compete with STFA for this energy, three significant ones can be ruled out: proton 
acceleration by STFA, Landau damping, and resistive dissipation of the turbulence.  The latter two have already been discussed.  Proton 
acceleration is a significant concern since Fermi acceleration of protons and heavy ions was in fact the very problem Fermi intended to 
solve.  Thermal protons in coronal flare plasmas are sub-Alfvenic and thus cannot meet the condition for 
mirroring \citep{LaRosa, Blackman3}.  However, \citet{MillerRoberts} argues convincingly that gyroresont interaction of protons and 
heavy ions with Alfv\'en waves can accelerate them to velocities above $v_A$ on a relatively short time scale.  Within their model, 
the ions then are accelerated by compressive magnetosonic waves at the expense of electron acceleration.  Recent 
RHESSI observations \citep{Hurford} indicate that the emission signatures of ions and electrons are spatially separated, 
with the ion emission associated with longer loops.  \citet{Miller04} concluded that these observations are consistent with the 
gyroresonance model of ion acceleration; as the loops grow longer, protons are more likely to reach super-Alfv\'enic speeds, and thus
can be accelerated by the magnetosonic waves.  This second phase of acceleration need not be gyroresonant.  While it appears
promising, further study is required to determine if STFA models can accomodate these results.

The strong dependence of the post-acceleration electron spectrum on the pitch
 angle scattering agent is both a positive and negative feature. It leaves STFA considerable
flexibility in matching various characteristics of solar flare X-rays which fall outside of the simple scenario studied in this
paper.  For example, \citet{Lin81} first observed a superhot component in a solar flare, which has since been supported by RHESSI 
observations \citet{Krucker03}.  This thermal, or nearly thermal, spectral component is seen at energies of up to $35$keV.  Within
our STFA framework, the superhot emission can easily be explained by an enhancement of pitch angle scattering at lower energies, 
either by increased whistler wave turbulence, or some other scattering agent.  While this flexibility naturally allows for the wide 
range of flare characteristics observed, it does not yet definitively solve the flare acceleration problem.  Instead, it shifts the 
focus exclusively to a well constrained, but largely unspecified, array of pitch angle scattering mechanisms.  
This is the single greatest obstacle to STFA models of acceleration.  
   
In short, STFA can naturally account for the thermal spectrum below
$10$keV,  and somewhat less naturally for
the non-thermal spectrum between $10$keV and $100$keV. There we have shown that
$\lambda_p$ must depend inversely on particle energy, in contrast
to that of pitch angle scattering by whistler waves below $\backsim 10$kev, which is proportional
to the particle energy.  Above $100$keV, shock acceleration is a natural
possibility; the needed injection of super-thermal electrons may be provided
by STFA operating at energies below $E_{br}$.
The knee at $100$keV remains the most difficult spectral feature to accommodate, and we have explained the difficult requirements to 
pitch angle scattering that this demands.  

E.B. thanks B. Chandran for discussions.  We thank J. Maron for sharing the results of his simulations and
acknowledge support from DOE grant DE-FG02-00ER54600 and the Laboratory for Laser Energetics at the University of Rochester.  RS 
acknowldges support from the DOE Horton Fellowship.  We also acknowledge the insightful critique and comments of the reviewer.

\section*{Appendix A: The Power Law Spectrum of Relativistic STFA}

Notice that the spectrum we obtain for STFA is different from the power law result of \citet{jones1994}.  This 
is a matter of regime; we discussed in the text the
acceleration of non-relativistic particles in a region of non-relativistic
turbulence, here we show 
that when the particles are relativistic, a power law spectrum emerges.

Recall that (\ref{Epm}) for fully 
relativistic electrons in a region of non-relativistic turbulence is given by (\ref{Estep})
\begin{eqnarray}
\nonumber
\delta E_{\pm R} = 2 E \left[\pm \frac{v_A v_\parallel}{c^2} + \frac{v^2_A}{c^2}\right],
\end{eqnarray}
where $E$ is the total energy, kinetic plus rest, of the electron before 
reflection.  Notice that if we the low velocity limit, $v \ll c$ where $E = mc^2$, the expression reduces to 
(\ref{Epm}).  
The relative velocity between the 
compression and electron for head-on and catch-up type 
interactions are still given by (\ref{erates})
so the steady acceleration rate is given by

\begin{equation}
\left(\frac{dE}{dt}\right)_S  = 2E \left[(R_+ - R_-)\frac{v_A v_\parallel}{c^2} + R\frac{v^2_A}{c^2}\right] 
			      = 4\frac{F}{\lambda}\frac{v^2_A}{c^2} v_\parallel E.
\label{rateHR}
\end{equation}
Alternatively, we can find the mean acceleration per reflection by multiplying equation \ref{rateHR} by $R^{-1}$

\begin{equation}
\left(\frac{dE}{dl}\right)_S = 4 \frac{v^2_A}{c^2} E.
\end{equation}  
In the highly relativistic limit, $E$ is just the kinetic energy, and we recover the familiar result \citep{jones1994}
that $dE/dt \propto E$.  This proportionality is expected to produce a power law.
We derive the power law spectrum for STFA of highly relativistic electrons by following the approach of
\citet{Bell} and assume that $p_{esc}$ is a constant in flare plasmas, independent of electron energy. 
We start by integrating $dE/dl$ to obtain $E(l)$

\begin{eqnarray}
l = \frac{1}{A}\ln{(E/E_0)}, \\ \nonumber
A= \frac{4 v^2_A}{c^2}.
\label{El}
\end{eqnarray}
The probability of an electron remaining in the acceleration region for at least $l$ reflections is given by

\begin{equation}
P(l+) = (1-p_{esc})^{l}.
\end{equation}
Taking the logarithm and substituting in for l from equation \ref{El}, gives

\begin{equation}
\ln{P(E+)} = l\ln{(1-p_{esc})} = \frac{1}{A}\ln{\left(\frac{E}{E_0}\right)}\ln{(1-p_{esc})}
	   = \ln{\left(\frac{E}{E_0}\right)^{(-p_{esc}/A) -1}}.
\end{equation}
where we used the approximation $ln{(1-p_{esc})} = -p_{esc}$ for $p_{esc} \ll 1$ in obtaining the 
expression to the right of the final equals sign.  Differentiating with 
respect to $E$, results in

\begin{equation}
P(E) = { E_0}\left(\frac{E}{E_0}\right)^{(-p_{esc}/A)-1},
\end{equation}
where $P(E)dE$ is the unnormalized probability of a post-acceleration electron having the energy $E$.   
In the limit, where $p_{esc}$ is extremely small, the relativistic STFA spectrum has power law index $\sim 1$.  In 
a plasma where $p_{esc} \sim A$, the power law index can grow larger, and the index is very sensitive to $p_{esc}$.   
In the third regime, where 
$p_{esc} \gg A$, electrons stream out of the turbulent volume quickly, do not experience much acceleration, and have a 
very steep power law energy distribution with virtually no very high energy electrons $(E \gg E_0)$.  

\section* {Appendix B: Derivation of Steady Acceleration Rate with $F_+ \neq F_-$}

In section 3.1 we derived the steady acceelration rate for electrons in a low $\beta$ turbulent magnetic plasma. This derivation was 
contingent on the assumption that $F_+ = F_- = F$, which is not strictly valid.  \citet{Blackman3} calculates $F$ for Fermi 
acceleration.  By resetting the limits of the integral in his eq (12), and renormalizing for the smaller phase space, one arrives 
at

\begin{equation}
F_{\pm} = \cos{\phi_m} \left[\pm \frac{v_A}{v} + \left(1-\left(\frac{v_A}{v}\right)^2(1-\cos^2\phi_m)\right)^{1/2}\right],
\label{fpm}
\end{equation}
where $\cos{\phi_m}$ is the minimum pitch angle at which an electron will reflect and $v_A/v$ is the ratio of 
the Alfv\'en speed to the electron speed.  We rename these quantities $A$ and $B$ respectively; both are small quantities.  
By taking a series expansion of eq (\ref{fpm}) and truncating it at second order in $B$, it can be simplified to
   
\begin{equation}
F_\pm = A\left[1 \pm B - \frac{1}{2} B^2 \right].
\end{equation}

Recall that from (\ref{erates}),

\begin{equation}
R_\pm = F_\pm \left(\frac{v_\parallel \pm v_A}{2\lambda}\right) = A\left[1 \pm B - \frac{1}{2} B^2\right](A \pm B) \frac{v}{2\lambda}.
\end{equation}
From this one easily obtains

\begin{equation}
R = (R_+ + R_-) = \frac{v}{\lambda} \left[ A^2\left(1 - \frac{1}{2} B^2\right) + AB^2 \right],
\end{equation}
and

\begin{equation}
(R_+ - R_-) = \frac{v}{\lambda} \left[A^2B + AB\left(1-\frac{1}{2}B^2\right)\right].
\end{equation}

This gives us all of the ingredients for calculating the steady acceleration from (\ref{sdefine})

\begin{eqnarray} \nonumber
\left(\frac {dE}{dt}\right)_S = 2m[(R_+ - R_-) v_\parallel v_A + (R_+ + R_-)v_A^2].
\end{eqnarray}
The resulting acceleration rate is

\begin{equation}
\left(\frac {dE}{dt}\right)_{Sb} = \frac{2mv^3}{\lambda} AB^2\left[2A + A^2 - \frac{1}{2}B - \frac{1}{2}AB^2 + B^2 \right], 
\label{sb}
\end{equation}
where we have added the additional subscript $b$ to indicate the distinction from the previously calculated rate.
The steady acceleration rate found in (\ref{ratecalcfirst}) from the assumption $F_+ = F_- = F$ is

\begin{equation}
\left(\frac{dE}{dt}\right)_S = \frac{4mv^3}{\lambda} A^2B^2.
\end{equation}
Note that provided $A > 4B$ this is the largest term in (\ref{sb}).  Indeed, for coronal flare plasma, $B \backsim 0.1$ at 
electron energy $E_0$ and decreases with increasing energy while $A \backsim 0.1$ as well at $E_0$, but is largely 
insensitive to eletron energy. At the onset of the power law regime, $E= 10keV = 30E_0$, and $B \backsim 0.01$; 
all terms of order $B^3$ 
or higher can be neglected, as can the term in $A^3$.  Thus we can safely use the assumption $F_+ = F_- = F$ in this 
regime, and (\ref{11}) is reasonable.

\begin{figure}
\plotone{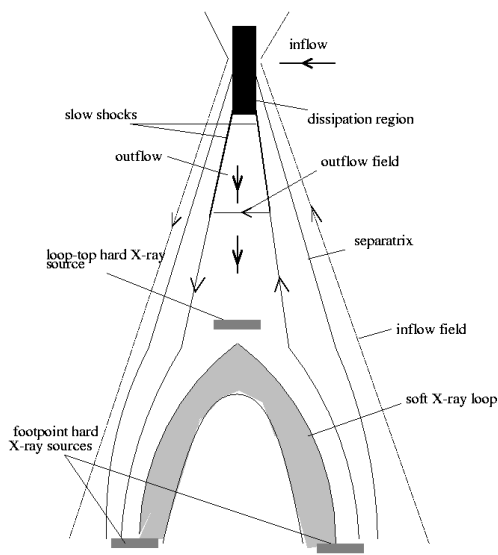}
\caption{From \citep{Blackman2}.  Sketch of a typical impulsive solar flare.  Note that the x-point 
reconnection occurs in the filled region at the top of the diagram.  Only the downward half of the outflow is 
shown.  Reproduced by permission of the AAS.}      
\label{flaresketch}
\end{figure}

\begin{figure}
\plotone{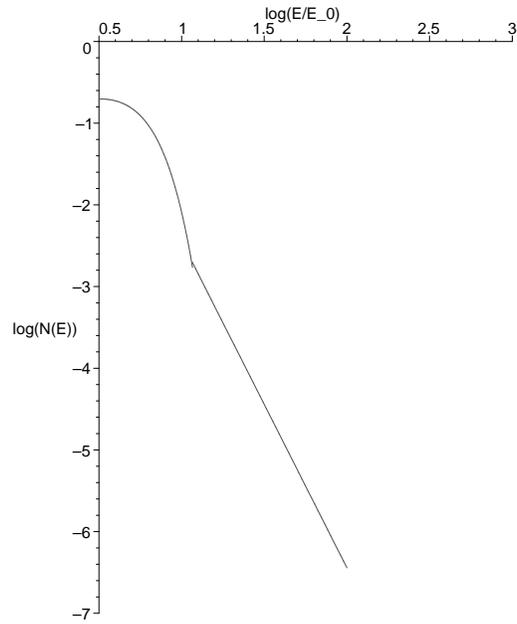} 
\caption{The spectrum of non-relativistic STFA under impulsive flare conditions with whistler wave turbulence and a second 
source of pitch angle scattering.  The
second pitch angle scattering source obeys the constraints required to produce a power law. $E_0 = 0.3$keV.}
\label{combspec}
\end{figure}

\end{document}